\documentclass[conference]{IEEEtran}
\IEEEoverridecommandlockouts
\usepackage{graphicx}
\usepackage{cite}
\usepackage{amsmath,amssymb,amsfonts}
\usepackage{algorithmic}
\usepackage{textcomp}
\usepackage{balance}
\usepackage{booktabs} 
\usepackage{subfigure}
\usepackage{flushend}
\usepackage{breqn}
\usepackage{enumerate}
\usepackage{tabulary}
\usepackage{multirow}
\def\BibTeX{{\rm B\kern-.05em{\sc i\kern-.025em b}\kern-.08em
    T\kern-.1667em\lower.7ex\hbox{E}\kern-.125emX}}
\begin{document}

\title{Cluster analysis of homicide rates in the Brazilian state of Goi\'as from 2002 to 2014 \\
}
\author{\IEEEauthorblockN{Samuel Bruno da Silva Sousa}
\IEEEauthorblockA{\textit{Institute of Science and Technology} \\
\textit{Federal University of S\~ao Paulo}\\
S\~ao Jos\'e dos Campos, SP - Brazil \\
samuel.ssousa@outlook.com}
\and
\IEEEauthorblockN{Ronaldo de Castro Del-Fiaco}
\IEEEauthorblockA{\textit{Campus Henrique Santillo} \\
\textit{State University of Goi\'as}\\
An\'apolis, GO - Brazil \\
ronaldo.delfiaco@ueg.br}
\and
\IEEEauthorblockN{Lilian Berton}
\IEEEauthorblockA{\textit{Institute of Science and Technology} \\
\textit{Federal University of S\~ao Paulo}\\
S\~ao Jos\'e dos Campos, SP - Brazil \\
lberton@unifesp.br}
}

\maketitle

\begin{abstract}
Homicide mortality is a worldwide concern and has occupied the agenda of researchers and public managers. In Brazil, homicide is the third leading cause of death in the general population and the first in the 15-39 age group. In South America, Brazil has the third highest homicide mortality, behind Venezuela and Colombia. To measure the impacts of violence it is important to assess health systems and criminal justice, as well as other areas. In this paper, we analyze the spatial distribution of homicide mortality in the state of Goi\'as, Center-West of Brazil, since the homicide rate increased from 24.5 per 100,000 in 2002 to 42.6 per 100,000 in 2014 in this location. Moreover, this state had the fifth position of homicides in Brazil in 2014. We considered socio-demographic variables for the state, performed analysis about correlation and employed three clustering algorithms: K-means, Density-based and Hierarchical. The results indicate the homicide rates are higher in cities neighbors of large urban centers, although these cities have the best socioeconomic indicators. 
\end{abstract}


\section{Introduction}
Violence appears in the 21st century as one of the major social problems in urban centers\cite{b20}. 
This phenomenon is noticed in several ways in different spaces. It is not unusual to see news from all around the world reporting violent acts every day. Both physical and psychological forms of violence may change the relationships among people and the space they live\cite{b20}. We notice violence in households, groups, communities, neighborhoods, cities, and so on. Homicide is one of the most frightful forms of violence, and its occurrences vary from a place to another. 

In the most part of the world, the homicide rates are low, excluding war zones. The homicide count in 2012\cite{b1} was around 437,000 victims in the world. In 2017, the global homicide rate per 100,000 inhabitants was 6.2. 78\% of the victims were male, and the countries with the highest homicide rates were Honduras (85.5), Venezuela (53.7), and Belize (44.7). On the other hand, Monaco, Liechtenstein, and Andorra did not register any case of homicide which made their rates per 100,000 inhabitants reach the value 0. Most of the countries on the top of the list of homicide rates\cite{b1} were located in Latin America or the Caribbean. These regions historically presented high homicide rates, which is related to drug trafficking, corruption in governments, and cartels operations\cite{b2}. 

In Brazil, the homicide rate per 100,000 inhabitants increased between 2002 to 2014. In 2002, the rate was 27.9 per 100,000, while in 2014 reached 29.7 per 100,000, according to data from Igarap\'e Institute\cite{b1}. The main reasons for this increase are related to drug trafficking, corruption, and lack of policing. Only in 2014, 59,681 people were murdered in Brazil. The country is also responsible for more than 10\% of the homicides in the world\cite{b2}. 

In some Brazilian states, the increase in homicide rates has been even higher. For instance, in the state of Goi\'as, the homicide rate increased from 24.5 per 100,000 in 2002 to 42.6 per 100,000 in 2014, totalizing 24,300 victims in the period\cite{b1}. The state of Goi\'as is located in the Center-West region of Brazil, surrounding the Brazilian capital, Bras\'ilia. Its population was estimated at around 6,7 million people in 2017\cite{b3}, and its larger city is the capital, Goi\^ania, with 1,4 million inhabitants. The state is formed by 246 municipalities whose social-economic features vary a lot from one to another. Although the state concentrated the fifth highest homicide rate amongst the Brazilian states in 2014, this theme is still under-researched. 

Brazil covers 8.5 million $km^2$ (47\% of South America), with an estimated population of 190 million inhabitants in 2010\cite{b15}.
The country is divided into 27 States with differing socioeconomic and health conditions.  Recently published studies in Brazil analyzed homicide and suicide mortality rates in different states\cite{b9, b10, b11, b12, b13}. Despite the beliefs that homicides cluster in specific regions\cite{b7}, no papers using machine learning techniques such as clustering algorithms to analyze homicide rates were found. Thus, this paper introduces a new approach in this field.

Organizing data into groupings is an important mode of understanding and learning. Cluster analysis can be defined as ``the formal study of methods and algorithms for grouping objects according to measured or perceived intrinsic characteristics or similarity, without prior knowledge of the number of clusters or any other information about their composition''\cite{b24}. Cluster analysis does not use category labels that tag objects with prior identifiers, such as class labels. The absence of category information distinguishes data clustering (unsupervised learning) from classification (supervised learning)\cite{b24}. 

The main contributions of this paper are: 1) evaluation of correlations between homicide and socio-demographic variables for the state of Goi\'as, Brazil; 2) employment of three clustering algorithms (K-means, Hierarchical and Density-based clustering) to cluster the cities and to identify critical areas for homicides; 3) pattern identification suggesting that, in Goi\'as, the homicide rates are higher in cities neighbors of large urban centers, like Bras\'ilia and Goi\^ania, despite these cities having the highest per capita income in the country. 

The rest of the paper is organized as follows: 
Section \ref{related_work} presents related work about homicide rate analysis in the world and in Brazil. Section \ref{materials} presents the materials and methods employed in this work. The objectives, datasets, and algorithms are described with details. Section \ref{results} presents the results of our analysis. Finally, Section \ref{conclusion} presents the final remarks and future works.

\section{Related works} \label{related_work}

A plenty of works has been developed on homicide rate analysis since the last half of the 20th century. The fields of sociology and criminology were the first to start researching this theme. The main purposes of those works were related to investigating whether demographic, economic, ecological, and social variables maintained some correlations to the variation in homicide rates across time and space\cite{b4}. Variables as resident racial segregation, racial inequality, extreme poverty, social capital, and unemployment rate were used for some well-succeeded findings\cite{b5}.

In the United States of America, gunshot violence is responsible for about 34,000 deaths annually\cite{b6}.  A paper published in 2017 used scan spatial statistics to analyze clusters of the gunshot occurrences within the city of Syracuse, New York. Amongst the results, it was noticed that the higher violence rate was related to environmental and economic disparities\cite{b6}.

In Central America, where the homicide rates are historically elevated, some works have been developed on the analysis of possible causes. In El Salvador, the clusters of homicides may be related to drug trafficking and organized crime\cite{b8}. And in Mexico, the spatial variation of homicides was explained as being linked to firearms possession, drug trafficking and social exclusion\cite{b14}.

Some papers analyzed mortality by homicide in Brazilian states. Souza et al.\cite{b9} takes an ecological approach to the situation of homicides in all the municipalities of Bahia for the male population aged 15 to 39, considering the health macro-region. Lozada et al.\cite{b10} analyzed the homicide mortality trend from 1979 to 2005 for males aged 15 to 49, living in the State of Paran\'a. Souza et al.\cite{b11} made an ecological study which analyzed the violence-related death records of women aged 10 years and older, in the Brazilian geographic regions, between 1980 and 2014. Bando and Lester\cite{b12} evaluated correlations between suicide, homicide and socio-demographic variables by an ecological study for the Brazilian states. Peres et al.\cite{b13} described homicide mortality in the municipality of S\~ao Paulo according to social-economic features of the victims and type of weapon.

As far as we know no paper studied the homicide rates focused on the Goi\'as state. Furthermore, we did not find any work using clustering algorithms to analyze the spatial distribution of homicides. 
The techniques and tools commonly used on this topic are: estimation of regression coefficients\cite{b4,b5}; spatial scan statistics\cite{b6}; SaTScan software methods\cite{b7}; Bayesian approach with Monte Carlo Markov Chain algorithm\cite{b8}; Moran's Global index\cite{b9}; descriptive statistics or correlation techniques\cite{b10,b12}; estimable functions and negative binomial regression\cite{b11}; SSPS software tools\cite{b13}; and multiple regression analysis (stepwise method)\cite{b14}. 

This paper aims to analyze spatial patterns of mortality by homicide for better understanding violence in the Goi\'as state and to identify risky areas. Due to the lack of use of clustering algorithms on this theme, this paper proposes the use of K-means, Hierarchical and Density-based algorithms for clustering, in an exploratory approach.


\section{Materials and methods} \label{materials}
This is an exploratory study of homicide mortality in the State of Goi\'as for the period from 2002 to 2014. This period was chosen in function of data availability. Goi\'as had an estimated population of 6,778,772 inhabitants in 2017 and has 246 municipalities, being recognized as the 12th most populated state in Brazil\cite{b3}. The following sections present the objectives (\ref{A1}), the  methodology (\ref{A2}), the datasets used (\ref{A3}) and the evaluation measure employed in the experiments (\ref{A4}). 

\subsection{Objectives}\label{A1}
The general objective of this study is to analyze the spatial distribution of mortality by homicide in the State of Goi\'as. The specific objectives can be divided as follows:
\begin{itemize}
\item Collect data about mortality from homicide rates (MHR) in Goi\'as from 2002 to 2014.
\item Collect demographic and socioeconomic characteristics of the various territorial areas for the purpose of correlated with homicide rates.
\item Apply cluster algorithms to group similar data, specifically, K-means, Hierarchical and Density-based algorithms.
\item Identify patterns in the clustering results to identify risky areas.
\end{itemize}

\subsection{Methodology}\label{A2}
The presence of clusters was evaluated by K-means, Hierarchical and Density-based algorithms.
The software used for building the database and for computing the clusters was R.\footnote{https://www.r-project.org/}

\textit{K-means clustering} aims to partition $n$ data instances into K clusters in which each instance belongs to the cluster with the nearest mean, serving as a cluster's prototype\cite{b21}. This is a very popular algorithm in data mining. It proceeds by alternating between two steps, given an initial set of K centroids $m_{1}$, $m_{2}$... $m_{n}$. The first step comprehends the assignment of each instance to the cluster with the least squared Euclidean Distance. In the second step, new means are calculated to be the centroids of the new clusters. During the iterative process, it is common that some instances skip from one cluster to another. The ending of the iterations occurs when the instances are established and, at this time, the K-means are truly the means of the groups formed\cite{b21}.

Let $\mu_k$ be the centroid of cluster $c_k$. The squared error between
$\mu_k$ and the points in cluster $c_k$ is defined as in Equation \ref{k_means}.

\begin{equation} \label{k_means}
G(c_k) = \sum_{\vec{x_i} \in c_k} ||\vec{x_i} - \mu_k||^2
\end{equation}

The goal of K-means is to minimize the sum of the squared error
overall K clusters, as defined by Equation \ref{K_means}.

\begin{equation} \label{K_means}
G(C) = \sum_{k=1}^{K} \sum_{x_i \in c_k} ||x_i - \mu_k||^2
\end{equation}

Automatically determining the number of clusters is a difficult problem in data clustering. K-means is run independently for different values of K and the partition that appears the most meaningful to the domain expert is selected. 

\textit{Hierarchical clustering} is a method of cluster analysis which seeks to build a hierarchy of clusters\cite{b22}. Its representation consists of a hierarchical system (dendrogram) like the taxonomic hierarchy in biology.  Strategies for hierarchical clustering generally fall into two types: 
\begin{itemize}
\item Agglomerative: This is a ``bottom up'' approach where each observation starts in its own cluster and pairs of clusters are merged as one moves up the hierarchy.
\item Divisive: This is a ``top-down'' approach where all observations start in one cluster, and splits are performed recursively as one moves down the hierarchy.
\end{itemize}

Formally, it is assumed that when there is a sequence of $n$ elements, represented by values from 1 to $n$, there is also a sequence of clusterings with length $m + 1$  ($C_0$, $C_1$, $C_2$, \ldots, $C_m$)\cite{b22}. Each cluster has a number $\alpha_i$ with its value. The cluster $C_0$ is considered the weakest clustering of $n$ elements ($\alpha_0$ = 0), and the cluster $C_m$ is considered the strongest clustering. The numbers $\alpha_i$ increase ($\alpha_i$ $\leq$ $\alpha_{i+1}$) as well as the clusters $C_i$ ($C_i$ $\leq$ $C_{i+1}$), which means each cluster $C_i$ is the merging (or union) of clusters $C_{i-1}$\cite{b22}.
In general, the merges and splits are determined in a greedy manner. The results of hierarchical clustering are usually presented in a dendrogram. The most common algorithms are Single-linkage and Complete-linkage clustering.

\textit{Density-based clustering} (DBSCAN) is one of the most common density-based clustering algorithms. It was designed to discover clusters and noise in a spatial dataset, requiring only one parameter for input whose value is determined by the user\cite{b23}. This algorithm works as follows: given a set of points in some space, it groups together points that are closely packed together (points with many nearby neighbors), marking as outliers points that lie alone in low-density regions (whose nearest neighbors are too far away).

The DBSCAN algorithm can be abstracted into the following steps:
\begin{itemize}
\item Find the $\epsilon$ neighbors of every point and identify the core points with more than a predefined number of neighbors.
\item Find the connected components of core points on the neighbor graph, ignoring all non-core points.
\item Assign each non-core point to a nearby cluster if the cluster is an $\epsilon$ neighbor, otherwise, signalize it as noise.
\end{itemize}

DBSCAN is an effective algorithm to discover clusters that present unusual shapes even in large spatial databases\cite{b23}.

\subsection{Dataset}\label{A3}
\begin{figure}[h!]
  \caption{Municipalities of the State of Goi\'as.} \label{goias}
  \centering
  \includegraphics[width=0.5\textwidth]{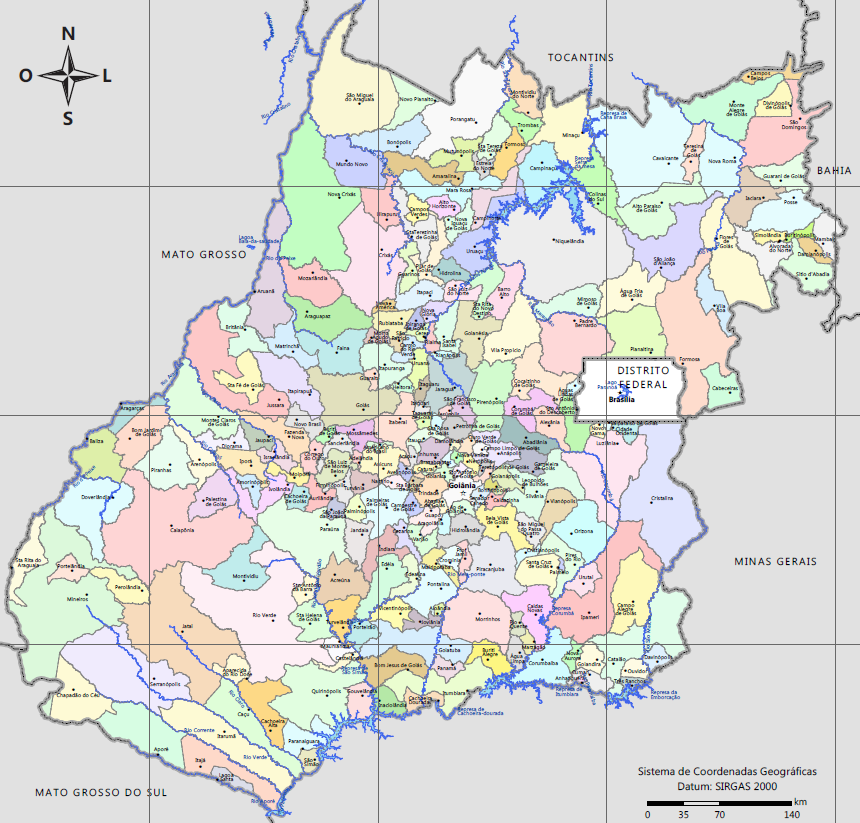}
  \text{Source: http://www.sieg.go.gov.br/rgg/atlas/index.html/.}
\end{figure}

Data from 246 municipalities of the state of Goi\'as (Figure \ref{goias}) were used as analysis units.

The data were collected between January and March of 2018 by downloads from official websites of organizations maintained by the Brazilian government and the United Nations. All data used in this analysis are publicly available for download on the Internet. They are highly reliable once they are published by governmental agencies and by internationally recognized organizations.  

The data on demographics consist of the population estimate for each municipality and its demographic density. They were extracted from the website of the Brazilian Institute of Geography and Statistics (IBGE)\cite{b3}. The values used in the query consisted of the municipalities' names.

The data on mortality were extracted from the Mortality Information System (SIM)\cite{b16}. This system is maintained and updated by the Brazilian Ministry of Health with data from mortality in general. In the query, we aimed to extract only the MHR, that included deaths by aggression, under the CID-10 codes X85 and Y09, and by legal intervention, under the codes Y35 and Y36. The selected values in the query were: the area of influence of the State of Goi\'as, the municipalities' names, the period in the extraction was between the years 2002 and 2014, and the codes of CID-10 mentioned above.

The data about the Basic Education Development Index (IDEB)\cite{b17} were extracted from the website of Instituto Nacional de Estudos e Pesquisas Educacionais An\'isio Teixeira (INEP), an agency from the Brazilian Ministry of Education. This index measures the quality of basic education in Brazil, according to international standards. The values for the index range from 0 to 10. In the query, we selected to consult by: municipality, public (federal, state or municipal) administrative dependency, series corresponding to the final years of Secondary School (8th and 9th years in Brazil).

The data on social and economic features were collected from the website of the United Nations Development Program (UNDP). In Brazil, UNDP develops an atlas of development in the country. This atlas, named Human Development Atlas in Brazil, covers all the Brazilian States and Federal District and its municipalities. In this study, we collected the data that comprehended the Municipal Human Development Index (MHDI)\cite{b18} and its variables, as life expectancy, education index, the percentage of total income appropriated by the 10\% richest, and Gini index. These data were available to be downloaded in a Microsoft Excel file.

Table \ref{variables} contains the name and description of all variables selected to compose the dataset. 

\begin{table}[h]
\caption{Description of the social-economic variables.}  \label{variables}
\vspace{0.1cm}
\begin{tabular}{l | l}
	\hline
	Variable & Description \\
	\hline
 	MHR & Total number of MHR since 2002 until 2014. \\
	POPULATION & Population counting in IBGE 2010's census. \\
	DEMOGDENSITY & Municipality population by its total area. \\
	IDEB2005 & Basic Education Development Index in 2005. \\
	IDEB2007 & Basic Education Development Index in 2007. \\
	IDEB2009 & Basic Education Development Index in 2009. \\
	IDEB2011 & Basic Education Development Index in 2011. \\
	IDEB2013 & Basic Education Development Index in 2013. \\
	LIFEEXPECT & Life expectation in 2010. \\
	GINI & Gini coefficient in 2010. \\
	INRICHEST10 & Rate of the overall income held by richest 10\%. \\
	EDUCLEVEL & Education level of adult population in 2010. \\
	MHDI & MHDI in 2010. \\
	MHDIE & MHDI (Education) in 2010. \\
	MHDIL & MHDI (Longevity) in 2010. \\
	MHDII & MHDI (Income) in 2010. \\  
    \hline
\end{tabular}
\end{table}

For this analysis, the dataset was preprocessed and analyzed in R. We have made use of Factoextra, Amap, FPC, Cluster, and DBSCAN packages. 

\subsection{Evaluation of results}\label{A4}
The process of evaluating the results obtained from a clustering algorithm is commonly called validation. There are two main types of grouping validation indexes: i) external indexes, which compare the group structure discovered with a previously known group structure; ii) internal indexes, which analyzes the structure of groups discovered in relation to some criterion, such as, for example, compactness or separability.

We employ the external index Silhouette, and the internal measures GAP statistic and Sum of Squares Within (SSW)  to validate the clustering results.
The Silhouette index (SIL) is calculated per data and the SIL of a group is the average of the SIL of all the data in the group. And the clustering SIL is the mean of the SIL of the groups. The higher the index value the better. It is described by Equation \ref{sil}.
\begin{equation} \label{sil}
SIL = \frac{a_i - b_i}{max\{a_i,b_i\}}
\end{equation}
where $a_i$ is the average distance from the data $i$ to all other data in its group and $b_i$ is the minimum distance of the data $i$ to all other data that do not belong to its group.

The GAP statistic compares the overall intra-cluster variation for a different number of clusters with their expected values under a null reference data distribution. The GAP statistic for a given value $k$ is defined as;
\begin{equation}\label{gap}
GAP_n(k) = E^{*}_{n} log(W_k) - log(w_k),
\end{equation}
where $E^{*}_{n}$ is an expectation in a sample of size $n$ from the reference distribution, obtained by bootstrapping by computing the average $log(W_k)$, and by generating $B$ copies of the dataset. The GAP statistic measures the deviation of the observed $W_k$ value from its expected value under a null hypothesis. 

SSW is a measure of compactness of a cluster structure, given by the distance between the centroid of a cluster and the instances assigned to it. The SSW equation has the form:
\begin{equation}\label{ssw}
SSW_M = \sum^{N}_{i = 1} {|| x_i - c_{Pi} ||}^2,
\end{equation}
where $x_i$ is an instance in each cluster whose centroid is $c_{Pi}$.

\section{Results and discussion} \label{results}
Section \ref{B1} presents the analysis of the variables and their correlation with homicide rate. Pearson's, Spearman's and Kendall's indexes were employed to measure correlation. Section \ref{B2} presents the cluster analysis. Algorithms Hierarchical, Density-based and K-means were employed to cluster the municipalities. Section \ref{B3} presents the discussion of the results.

\subsection{Social-economic variables analysis} \label{B1}
\begin{figure}[h!]
  \caption{MHR in the State of Goi\'as by year.} \label{MHR}
  \centering
  \includegraphics[width=0.50\textwidth]{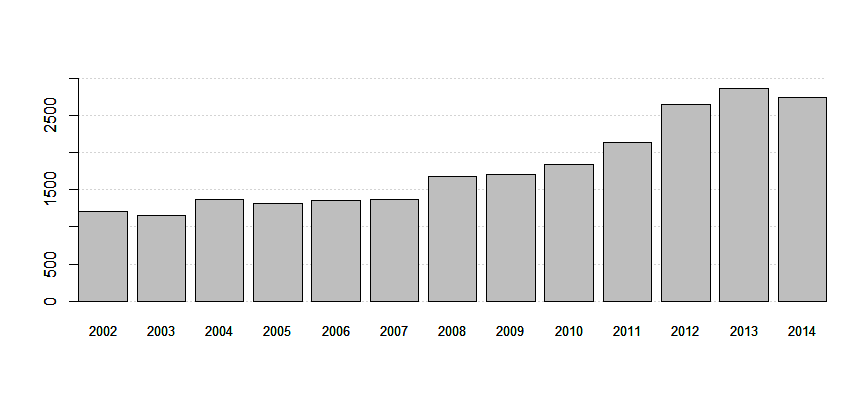}
\end{figure}

From 2002 to 2014, 24,300 people were murdered in the State of Goi\'as\cite{b1}. The number of MHR did not vary a lot between 2002 and 2007, but from 2007 to 2013 it increased year-to-year and reached the highest record in 2013 (Figure \ref{MHR}). In 2014, it was noticed a slight decrease in the number of MHR, but this was still the second highest record for the period. 

\begin{figure}[h!]
  \caption{Distribution of Gini coefficient in Goi\'as (2010).} \label{GINI}
  \centering
  \includegraphics[width=0.50\textwidth]{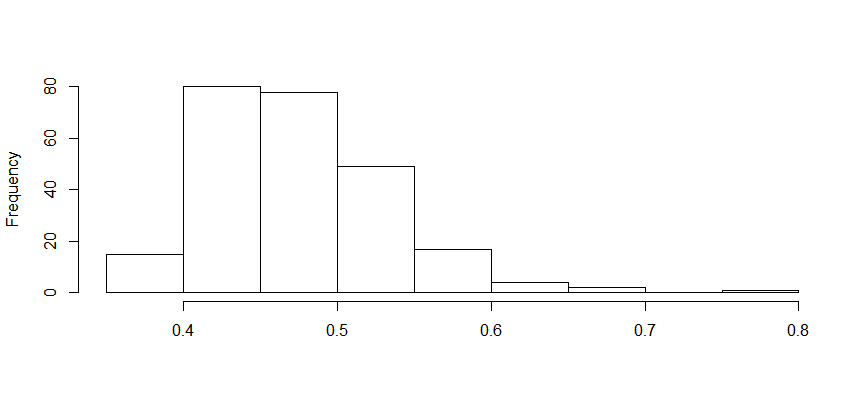}
\end{figure}

The Gini coefficient provides a measure of income distribution among individuals or households within a territory, such as a country, a state, a municipality, etc. A value 0 represents perfect equality, and a value 100 represents a total inequality. In our analysis, we used a normalized range of Gini coefficient from 0 to 1, in the way it was presented on the data source. Figure \ref{GINI} presents the distribution of Gini coefficient among the municipalities in the State of Goi\'as. Fewer than 20 municipalities have presented values for Gini index under 0.4. Thus, it is possible to notice that the State of Goi\'as presents high-income disparities among its municipalities and within most of the municipalities too. For instance, the state's largest city, Goi\^ania, presented a Gini value of 0.58. 

\begin{figure}[h!]
  \caption{Distribution of education level of adult population (2010).} \label{Education}
  \centering
  \includegraphics[width=0.50\textwidth]{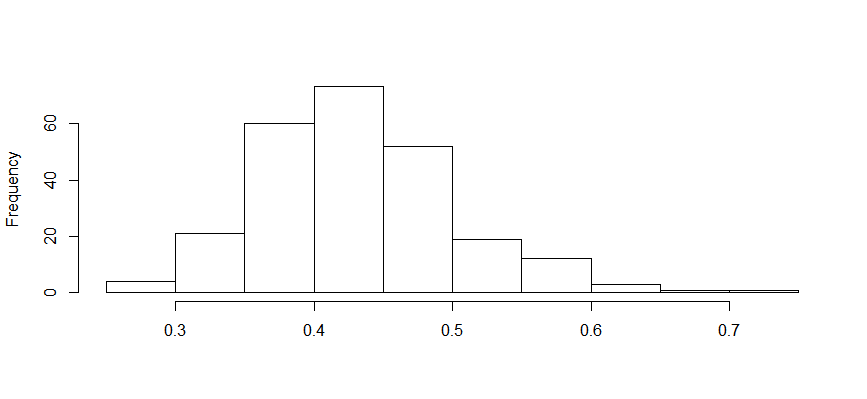}
\end{figure}

The educational variable EDUCLEVEL is related to the percent of the adult population that has completed the last year of Secondary School. In most of the municipalities, fewer than 60\% of the adult population completed Secondary School (Figure \ref{Education}). This situation is due to the fact of the most of the population lived in rural areas until the decade of 1970\cite{b19}, when schools and universities were concentrated in the largest cities. So the rural population was aside from formal education.

\begin{figure}[h!]
  \caption{IDEB variation through the years.} \label{ideb}
  \centering
  \includegraphics[width=0.50\textwidth]{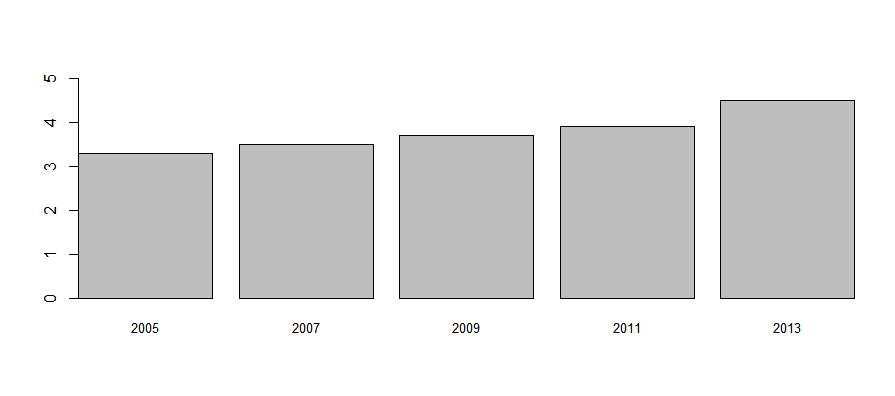}
\end{figure}

The variable IDEB consists of the Basic Education Development Index. In this study, we analyzed the values for IDEB in the last years of Secondary School for public schools. The means for IDEB from 2003 to 2013, are shown in Figure \ref{ideb}. It is possible to notice that the values of IDEB were increasing through the years, what means improvements in the quality of public basic education. In 2013, the IDEB mean for the state reached 4.5. In this year, the municipalities with the highest IDEB means were Perol\^andia (6.2), C\'orrego do Ouro (5.9), and Ivol\^andia (5.9), that are all small cities. On the other hand, the cities with the lowest IDEB means in 2013 were \'Aguas Lindas de Goi\'as (3.4), \'Agua Fria de Goi\'as (3.5), and Cavalcante (3.6), where the first two are located in the Surroundings of Federal District. 

\begin{figure}[h!]
  \caption{Distribution of Municipal Human Development Index in Goi\'as (2010).} \label{MHDI}
  \centering
  \includegraphics[width=0.50\textwidth]{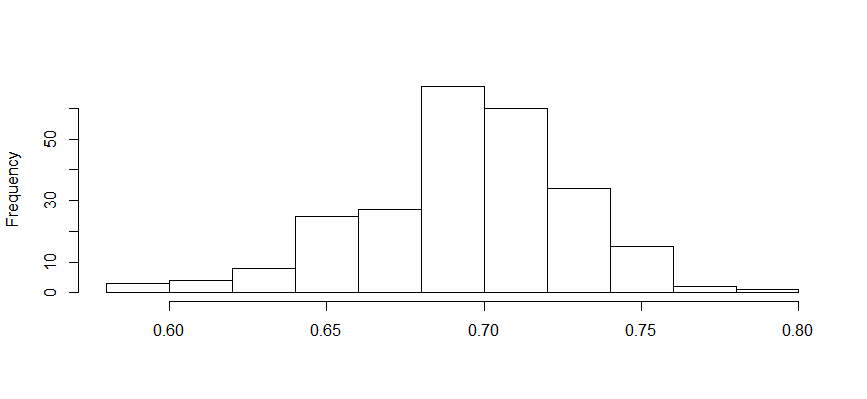}
\end{figure}

In 2012 UNDP started calculating MHDI for all municipalities in Brazil. This index is calculated the same way as HDI, which considers three dimensions (income, education, and longevity). Despite having the seventh highest HDI in Brazil (0.735) in 2010\cite{b3}, the State of Goi\'as presents a large disparity in the values of HDI among its municipalities (Figure \ref{MHDI}). More than half of the municipalities of Goi\'as have values for MHDI under 0.70, i.e., medium and low human development.  

\begin{figure*}[h!]
\caption{Regression lines between socio-economic variables and MHR (Color online).}
\centering
\subfigure[Population]{
    \centering
    \includegraphics[width=0.315\hsize]{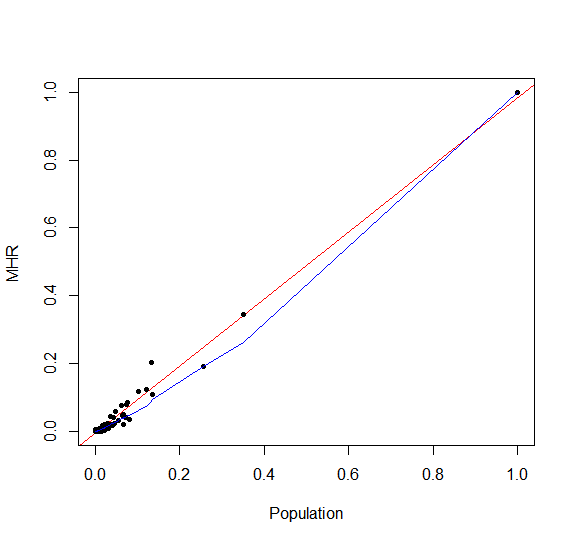}
    \label{fig:m1}}
    \hfil
\subfigure[Demographic density]{
    \centering
    \includegraphics[width=0.315\hsize]{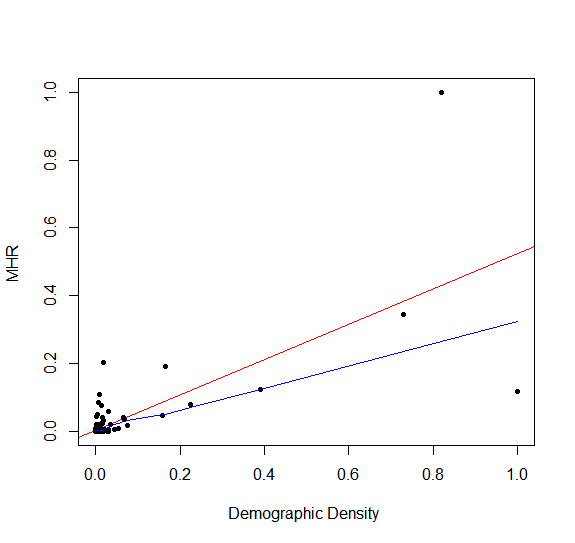}
    \label{fig:m2}}
        \hfil
\subfigure[Life expectation]{
    \centering
    \includegraphics[width=0.315\hsize]{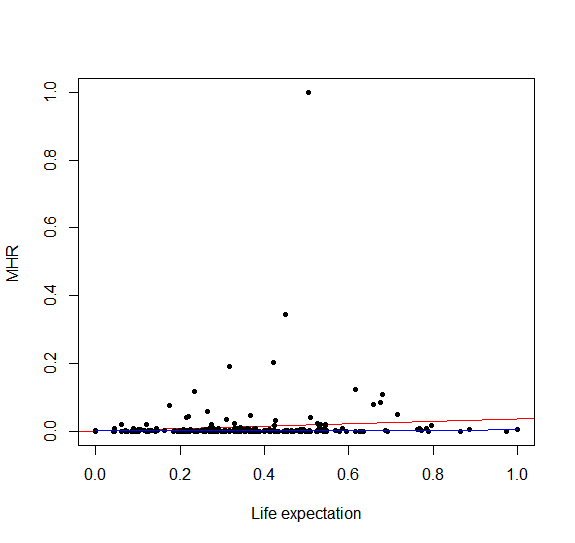}
    \label{fig:m3}}
    \vspace{0.1mm}
\subfigure[IDEB index]{
    \centering
    \includegraphics[width=0.315\hsize]{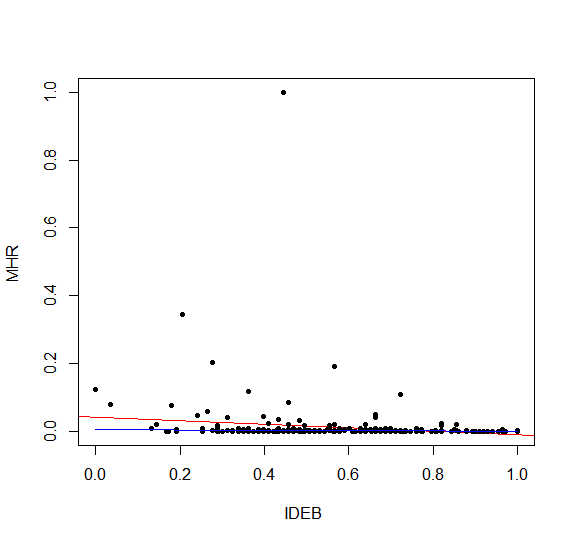}
    \label{fig:m4}}
\hfil
\subfigure[Educational level]{
    \centering
    \includegraphics[width=0.315\hsize]{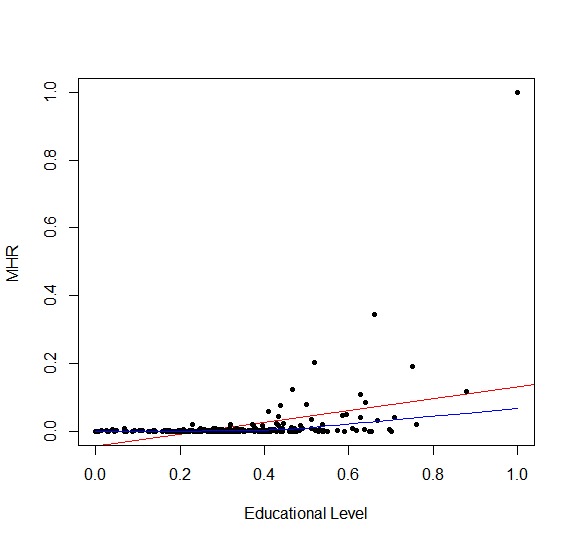}
    \label{fig:m5}}
\hfil
\subfigure[Gini index]{
    \centering
    \includegraphics[width=0.315\hsize]{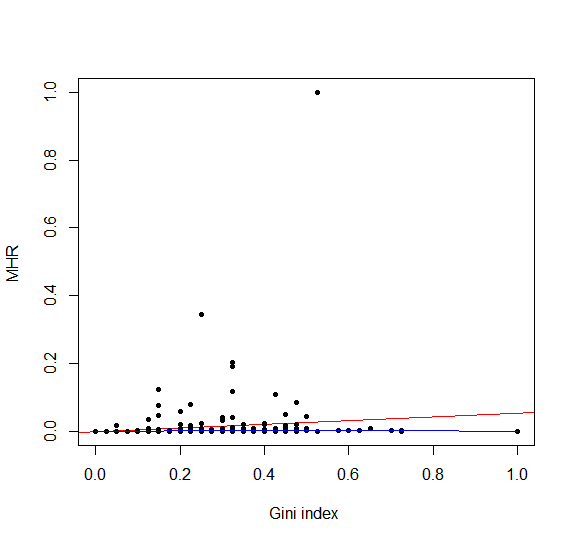}
    \label{fig:m6}}
\vspace{0.1mm}
\subfigure[Income held by richest 10\%]{
    \centering
    \includegraphics[width=0.315\hsize]{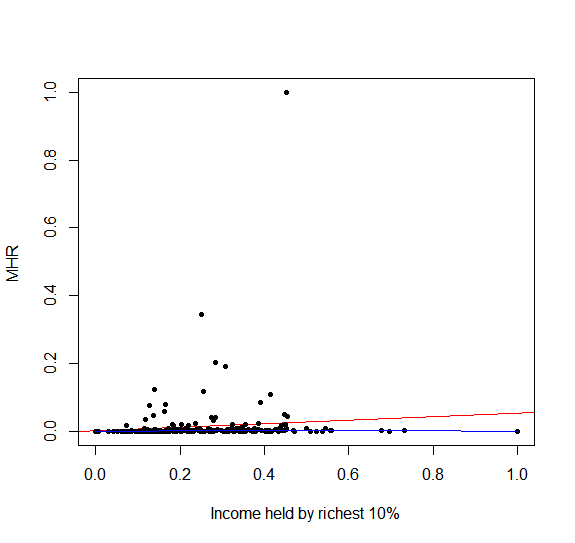}
    \label{fig:m7}}
\hfil
\subfigure[MHDI]{
    \centering
    \includegraphics[width=0.315\hsize]{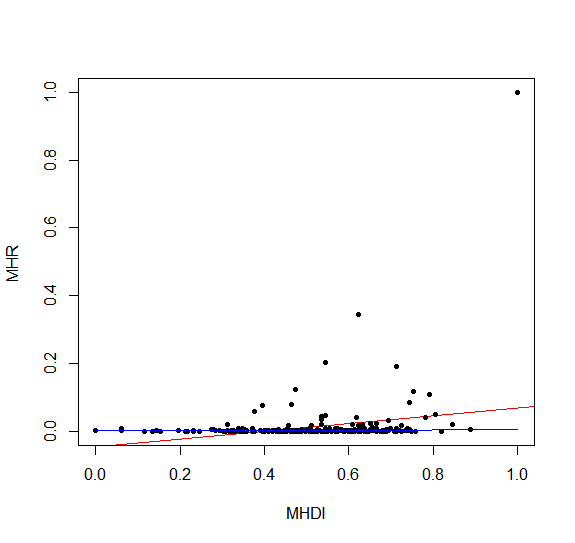}
    \label{fig:m8}}
\hfil
\subfigure[MHDI (Education)]{
    \centering
    \includegraphics[width=0.315\hsize]{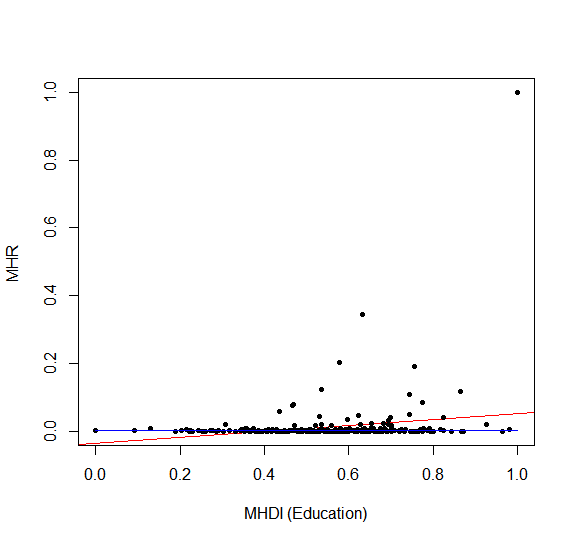}
    \label{fig:m9}}
\label{lines}
\end{figure*}

We also have calculated simple linear regression and local regression (LOWESS) between the dataset variables. The results are shown in Figure \ref{lines}. Simple linear regression is a statistical approach to model a relationship in the values of a dependent and one independent variable. This approach is commonly used in forecasting, error reduction, and relationships explanation. LOWESS is a non-parametric approach which makes use of a multiple linear method in a model based in k-Nearest Neighbor, popularly employed in trends forecasting of variables relationship. 
 
 The variables population (Sub-figure 7(a)) and demographic density (Sub-figure 7(b)) presented strong relationships with MHR values. In these sub-figures, it is possible to notice that both simple regression lines (red lines) and LOWESS lines (blue lines) assume rising values. Otherwise, the other variables are shown in the Sub-figures 7(c) to 7(i) presented weak relationships with MHR values. Both regression lines in these sub-figures assumed values equal or close to 0. The lines for the variables MHDI (Longevity) and MHDI (Income) also presented the values of simple and LOWESS regression close to 0.

Table \ref{Correlation} presents the Pearson's, Spearman's and Kendall's coefficient correlations between the social-economic variables considered and the MHR. The values of the correlations are between -1 and +1. The signal indicates the direction, whether the correlation is positive or negative, and the absolute value of the variable indicates the strength of the correlation. Values between 0.7 and 0.9 indicate a strong correlation. Notice that Population has a strong correlation with MHR as indicated by the three coefficients. Demographic Density also has a strong correlation, as indicated by the Pearson's coefficient. 

\begin{table}[h]
\centering
\caption{Correlations between the total of MHR and social-economic variables.} \label{Correlation}
\vspace{0.1cm}
\begin{tabular}{l | l | l | l}
	\hline
	Variable & Pearson's c. & Spearman's c. & Kendall's c.\\
    \hline
    POPULATION		&		\textbf{0.9915637}	&		\textbf{0.8932006}	& 	\textbf{0.7385185} 	\\
	DEMOGDENSITY	&		\textbf{0.7262442}	&		0.3446707	& 	0.2389337	\\
	IDEB		&		-0.1440371	&		-0.2019128	& 	-0.1417871	\\
	LIFEEXPECT			&		0.09612802	&		0.1669409	& 	0.1119476 	\\
	GINI			&		0.1134491	&		0.2435379	& 	0.1748277	\\
	INRICHEST10		&		0.1039698	&		0.2740375	& 	0.1939559	\\
	EDUCLEVEL	&		0.3994967	&		0.4042891	& 	0.287134	\\
	MHDI 			&		0.248962	&		0.1948997	& 	0.133404	\\
	MHDIE			&		0.206129	&		0.08632882	& 	0.05902461 	\\
	MHDIL			&		0.096134	&		0.1656033	& 	0.1120611	\\
	MHDII			&		0.2649425	&		0.2717906	& 	0.1852401	\\
	\hline    
 \end{tabular}
 \end{table}

\begin{figure}[!h]
\caption{Dataset distance matrix (Color online).} \label{distance}
\centering
\includegraphics[width=0.49\textwidth]{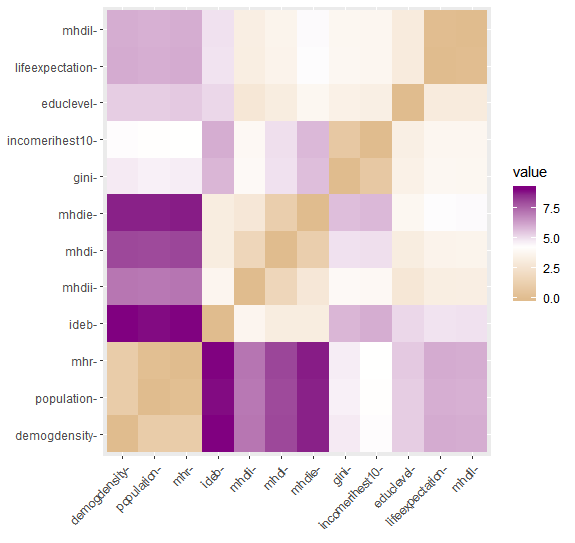}
\end{figure}

Before applying the cluster algorithms, we computed the distance between the dataset columns (Figure \ref{distance}), to visualize the (dis)similarity within the dataset. It is possible to notice that values for demographic density, population, and MHR are close to each other. At the same time, these variables present values far from the educational, economic, and MHDI variables values, showing to us that in the state of Goi\'as population growth was not accompanied by educational and human development growth.

\subsection{Cluster analysis} \label{B2}
This section presents the results of applying the Hierarchical, Density-based and K-means algorithms in the municipalities of Goi\'as. We validate the clustering results with the Silhouette measure, GAP statistic, and SSW. 

\subsubsection{Hierarchical clustering}
\begin{figure*}[h!]
\caption{Results of Agglomerative Hierarchical Algorithm (Color online).}\label{hclusters}
\includegraphics[width=\linewidth]{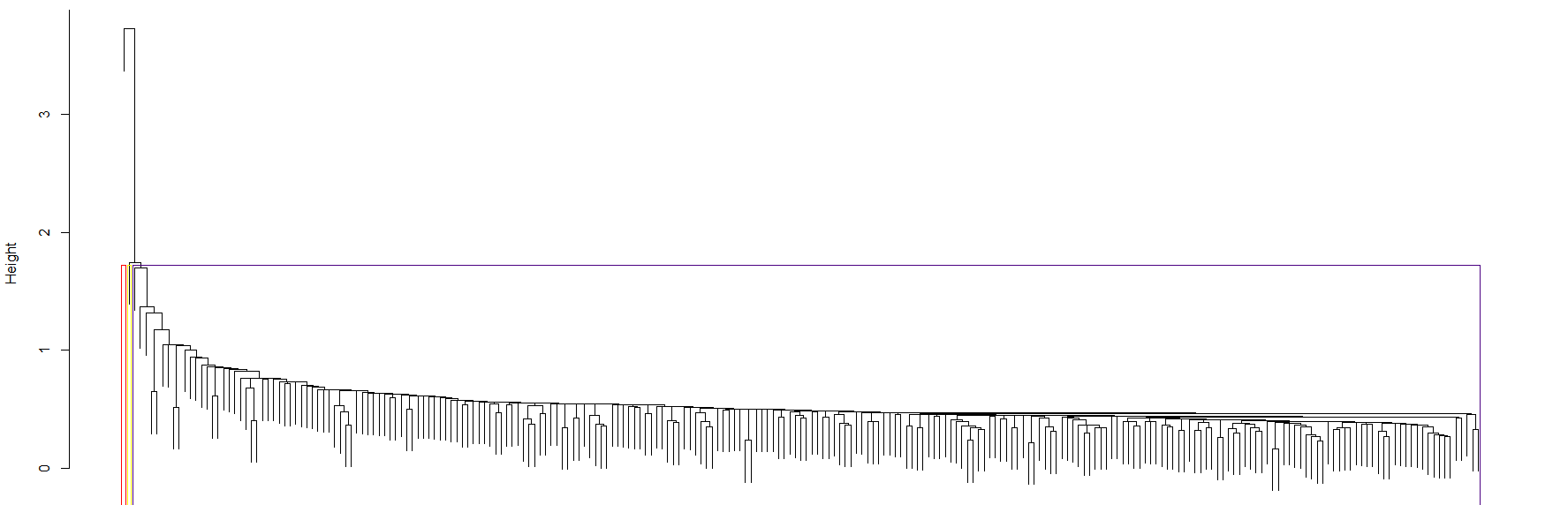}
\end{figure*}

The hierarchical clustering algorithm, with single linkage method and Manhattan distance, generated clusters with few municipalities. We have cut the dendrogram into 3 partitions, which is shown in the Figure \ref{hclusters}. The optimal number of clusters to cut the tree was decided by the results of Silhouette index and GAP statistic. We used the value which presented the biggest difference to its successor, and the optimum GAP statistic is given by $GAP_{(k)} > GAP_{(k+1)} - S_{k+1}$, being $S_{k+1}$ the standard deviation multiplied by the square of 1 plus the inverse of B copies of the reference dataset done by bootstrapping.

This algorithm assigned to the first cluster (in red) the largest city of the state, Goi\^ania; assigned to the second cluster (in yellow) the city with the highest demographic density, Valpara\'iso de Goi\'as; and assigned to the third cluster (in purple) the other cities in the state, despite of the dissimilarity between them.

\subsubsection{Density-based clustering}
\begin{figure}[h!]
\caption{Results of DBSCAN (Color online).} \label{dbclusters}
\centering
\includegraphics[width=0.51\textwidth]{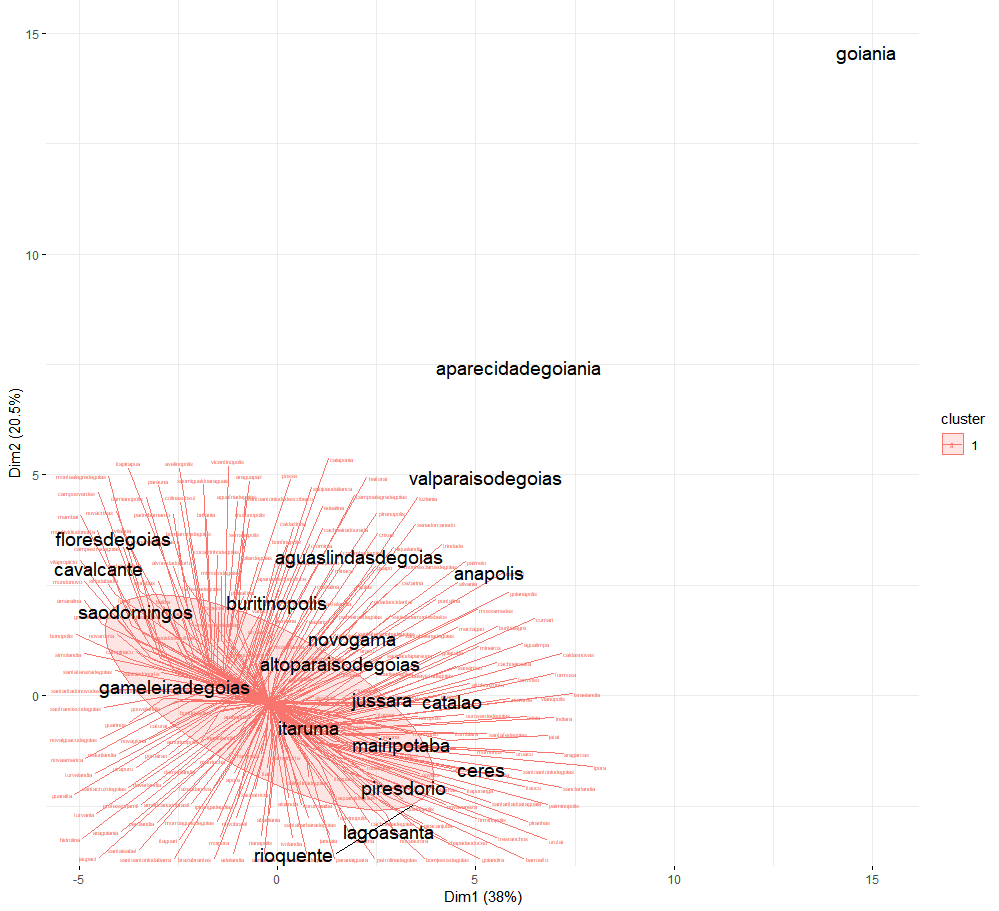}
\end{figure}

The DBSCAN algorithm generates just one cluster and assigned as outliers 20 municipalities, amongst these outliers, there are: the largest cities of the state, the municipality with the highest demographic density, and small cities whose values for MHR are 0 or close to it. This algorithm did not generate a meaningful result, because of the dissimilarity between the instances of the dataset. Most of the municipalities assigned as outliers hold the highest values for the variables population, demographic density and MHR. The results of DBSCAN are shown in the Figure \ref{dbclusters}.

\begin{figure}[h!]
\caption{Results of K-means clustering with K = 4 (Color online).} \label{kclusters}
\centering
\includegraphics[width=0.5\textwidth]{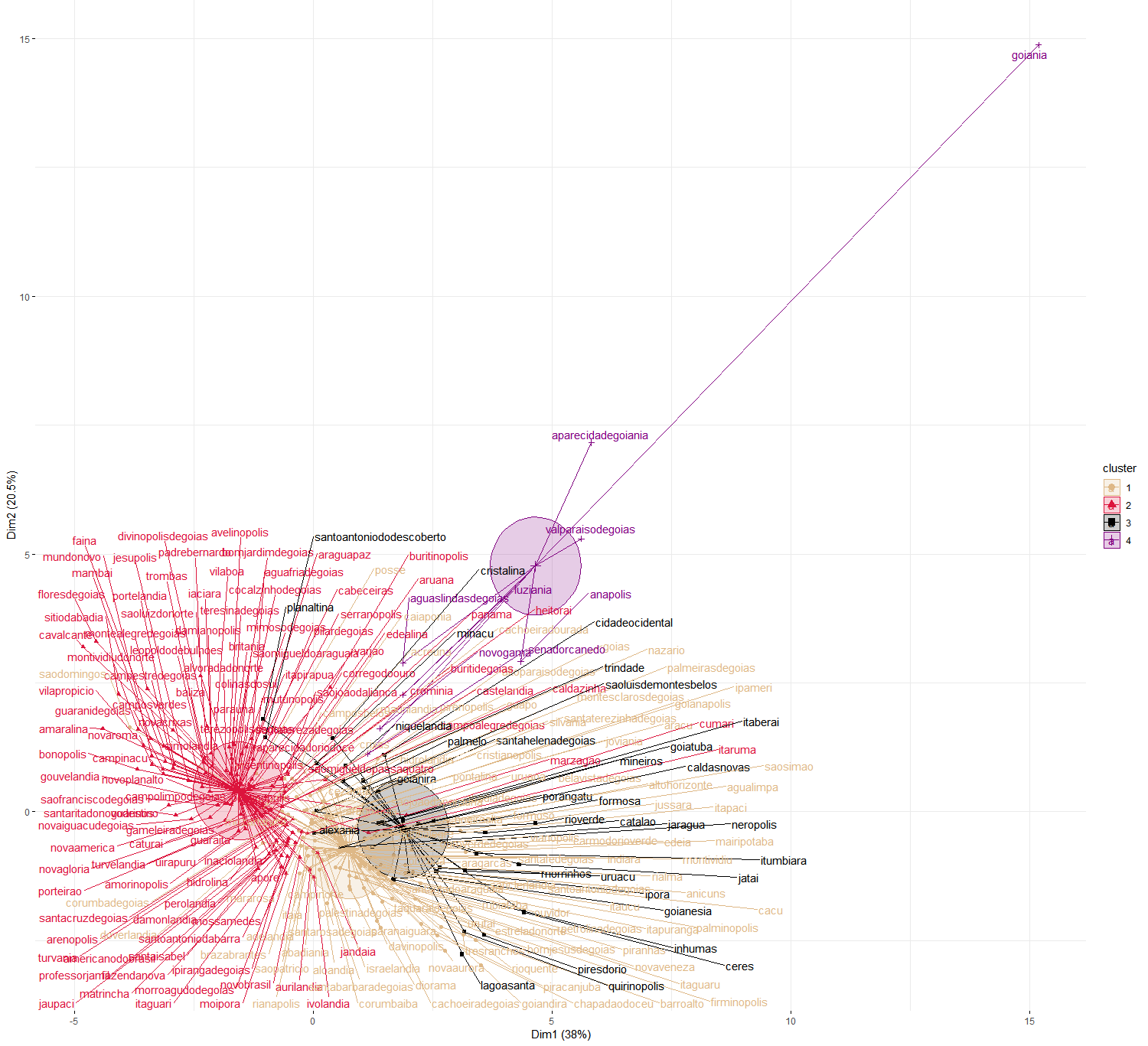}
\end{figure} 

\subsubsection{K-means clustering}
We employed different K values (from 1 to 5) in the K-means clustering algorithm. We also employed different distance measures, such as Euclidean, Manhattan, Pearson, and Canberra. The one that provided the most meaningful set of clusters was Canberra, once this measure is not sensible to outliers presence. It was decided to start K with 5 because the state has 5 regions and maybe this should influence the distribution of MHR. With K = 2, 3, and 5, the results were not interesting. In every application, the algorithm assigned Goi\^ania to the same cluster as small cities. The K = 4 value generated groups more interesting as shown in Figure \ref{kclusters}. The results signal different dynamics in the distribution of MHR among the municipalities of Goi\'as. The results of K-means algorithm are shown in Figure \ref{kclusters}.

In cluster 1 (beige), there are 96 municipalities predominantly small cities. The cluster presented the second lowest mean for the number of MHR and variables, such as population, demographic density, life expectation, Gini index, educational level of adult population, MHDI and its tree dimensions (Education, Longevity, and Income). Otherwise, this cluster has the second highest mean for income held by the richest 10\% and the highest mean for IDEB index amongst all the clusters.  

In cluster 2 (pink), there are 108 municipalities, which are also small cities close to the lowest means for population, demographic density, income held by the richest 10\%, educational level of adult population, MHDI and its three dimensions and MHR amongst all the clusters. Furthermore, this cluster has the second lowest IDEB and Gini index means. As the cities in this cluster are predominantly small, far from the two metropolis in the region, the educational level of the adult population tends to be low because, as cited in Sec. \ref{B1}, most of the state population was aside from formal education until the decade of 1970.

In cluster 3 (black), there are 33 municipalities mid-sized  like Rio Verde, Cidade Ocidental, Catal\~ao, Trindade, and Itumbiara; and some small cities. This cluster presented the highest means of life expectation, Gini index, income held by the richest 10\%, MHDI, MHDI (Longevity), and MHDI (Income). Furthermore, it was noticed in this cluster the second highest means for the variables population, demographic density, educational level, MHDI (Education), IDEB, and MHR.
 
The cluster 4 (purple) aggregates only 8 municipalities, all of them big or mid-sized cities. In this cluster, there are the municipalities of Goi\^ania, Aparecida de Goi\^ania, An\'apolis, Valpara\'iso de Goi\'as, \'Aguas Lindas de Goi\'as, Senador Canedo, Novo Gama, and Luzi\^ania. These cities compose the region between the capital of the state of Goi\'as, Goi\^ania, and the capital of Brazil, Bras\'ilia. The means of the population, demographic density, educational level, MHDI (Education), and MHR were the highest amongst the 4 clusters. At the same time, the cluster presented the second highest means for life expectation, Gini index, MHDI, MHDI (Longevity), and MHDI (Income). The mean income held by the richest 10\% was the third highest. Although having the highest value for the educational level of the adult population, the index which measures the quality of basic education -- IDEB -- presented the lowest mean amongst all the clusters. Furthermore, this cluster held 59.59\% of the MHR in the state from 2002 until 2014.

\begin{figure}[h!]
  \caption{Results of K-means clustering.} \label{goias-groups}
  \centering
  \includegraphics[width=0.5\textwidth]{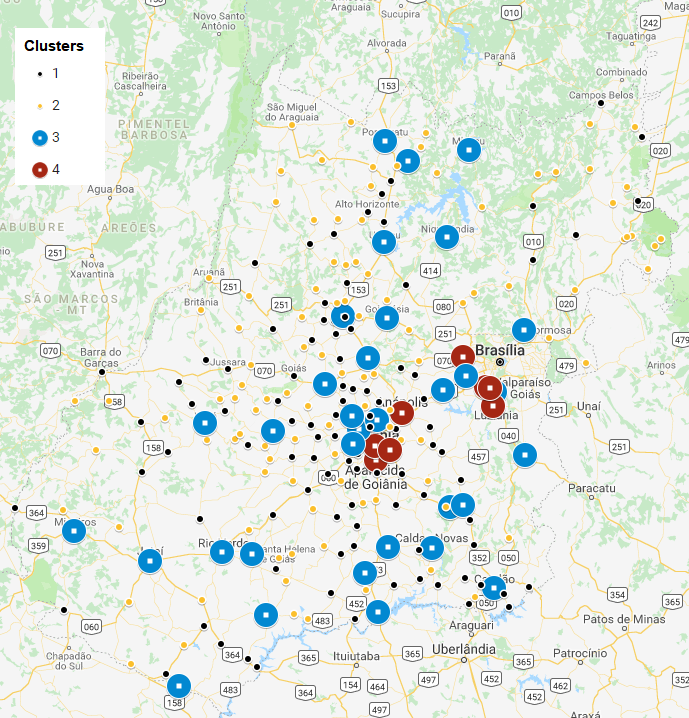}
\end{figure}

In the Fig. \ref{goias-groups}, we plotted the results of K-means clustering. The small black points represent the cluster 1 aforementioned. The yellow points represent the cluster 2. The blue points represent the cluster 3, and the big red points represent the cluster 4. It is possible to see that all the cities within the fourth cluster are located close to the region largest cities, Goi\^ania and Bras\'ilia, and Goi\^ania is also in this cluster.

\subsection{Discussion} \label{B3} 

Applying clustering algorithms it was possible to identify clusters of homicides in the state that are not merely random. Such clusters were considered as critical areas for homicides. It was also possible to notice different dynamics in the distribution of MHR in the state. The largest cities presented most of the homicide cases, despite its income and human development levels being higher than most of the other municipalities. On the other hand, these large cities present low values for IDEB index, suggesting that although having most people completing Elementary and Secondary School degrees, the quality of teaching in public schools is still low.

We also computed Silhouette measure, GAP statistic, and SSW to validate the clustering results (Table \ref{Validation}).
The values in bold represent the optimum values for these measures. The optimum Silhouette value for each algorithm was assumed as the value with the biggest difference to its successor in the table. Optimum GAP statistic values were determined by selecting the lowest which satisfied the condition: $GAP_{(k)} > GAP_{(k+1)} - S_{k+1}$. The optimal values for SSW were assumed by analyzing the bend (knee) in the curves of SSW, i.e., the SSW values in bold were the bends (knees) in their respective curves of SSW.

\begin{table}[h!]
\centering
\caption{Validation measures.} \label{Validation}
\vspace{0.1cm}
\begin{tabular}{l | l | l | l }
	\hline
	Clustering algorithm & Silhouette & GAP & SSW\\
    \hline
    K-means (2)	&	0.97 & 0.717 & 22.55 \\
	K-means (3)	&	0.89 & 0.748 &  26.10 \\
	K-means (4)	&	\textbf{0.84} & \textbf{0.765} & \textbf{23.96} \\
	K-means (5)	&	0.72 & 0.769 & 22.44\\
    Hierarchical (2) & 0.97	& 0.722 & 50.00 \\
    Hierarchical (3) & \textbf{0.94} & \textbf{0.742} & 44.30 \\
    Hierarchical (4) & 0.86	& 0.731 & 40.00 \\
    Hierarchical (5) & 0.85	& 0.743 & \textbf{36.80} \\
    Density-based & \textbf{0.97} & \textbf{0.705} & \textbf{66.71}\\
	\hline    
 \end{tabular}
 \end{table}

Measures such as Silhouette, GAP statistic, and SSW helps us to measure the clustering quality under a mathematical perspective. As clustering is an unsupervised task of data mining and Machine Learning, there is no label to be predicted/classified, and no training and test phases, to lately measure the error rate. So a meaningful clustering result is that one whose interpretation makes sense for a human being \cite{b25}.

Hierarchical and K-means clustering generates groups separating the large cities from small ones. The Silhouette measure indicates high similarity between the instances clustered. However, when the number of clusters was increased, the Silhouette measure decreased. The value 4 for K in K-means clustering was assumed as optimal because of GAP statistic value for K = 4 is bigger than the value for K = 5, i.e., $GAP_{(k=4)} > GAP_{(k=5)} - S_{k=5}$. 

K-means with K = 2 value achieves the highest Silhouette and the lowest SSW, but this clustering generates one group with 147 small cities and another group with 99 large cities. However, K = 4 gives a more interesting separation splitting the outliers cities, and K = 5 presents a low value for Silhouette, which suggests a weak clustering.

Density-based clustering resulted in just one cluster, and assigned as outliers 20 municipalities, amongst them the state largest cities. The Silhouette value for this results was also high, as well as the values for GAP statistic and SSW. 

Few studies bring information about all the municipalities from the State of Goi\'as, especially in the interior regions. Our contribution emphasizes the importance of the use of spatial analysis to understand violence indicators and geographic distribution.

\section{Conclusion} \label{conclusion}
In this paper, we employed three clustering algorithms: Hierarchical, Density-based and K-means to analyze spatial patterns of mortality by homicide in the Brazilian state of Goi\'as. 

The state's three largest cities, Goi\^ania, Aparecida de Goi\^ania, and An\'apolis, had the highest numbers of MHR. The municipalities in the region known as Surroundings of the Federal District also had elevated numbers of MHR. In the other municipalities, the numbers were low. All clustering algorithms separated these regions with high MHR from the others. However, K-means with K = 4 results in a better separation.

The means for population, Gini index, MHDI, IDEB, income concentration, and educational level of adult population varied among the municipalities as well as the number of MHR. The highest IDEB values were found in the smallest cities, which presented low homicide rates. In general, we noticed three scenarios for the occurrence of MHR in the state of Goi\'as. The first is characterized by a  low occurrence of MHR in the small cities, as shown by the clusters 1 and 2 in the Subsection \ref{B2}. The second scenario refers to the occurrence of MHR in mid-sized cities in municipalities with low demographic density and high-income concentration, as shown by the cluster 3. The last scenario is related to the occurrence of MHR in the state's municipalities with highest values of demographic density, low values of IDEB index, although having high MHDI and income, and comprising the metropolitan areas of Goi\^ania and Bras\'ilia, as noticed in the cluster 4. 

The use of clustering algorithms in the analysis of the distribution of MHR was an experimental approach, once we did not find studies that did it. In general, the behavior of the algorithms was satisfactory. Hierarchical clustering and DBSCAM provided results influenced principally by the population size, although K-means, with K = 4, showed an interesting separation whose quality was confirmed by its Silhouette measure, GAP statistic, and SSW.

We expect this study to be useful to improve the knowledge about the distribution of MHR in the State of Goi\'as and to aid the state's government to define risky regions, where policing should be more effective. This study is also a source to know the variation of social-economic variables among the state's municipalities.

As future work other analysis employing different variables to explain the rates of homicides in certain regions could be done, like precarious socioeconomic conditions, social inequality, etc. 

\section{Acknolowdgments} 
This work was conducted during a scholarship at ICT-UNIFESP -- Institute of Science and Technology of Federal University of S\~ao Paulo. Financed by CAPES -- Brazilian Federal Agency for Support and Evaluation of Graduate Education within the Ministry of Education of Brazil.

\balance

\end{document}